\documentclass{iau} 
\usepackage{graphicx}
\usepackage{natbib}
\usepackage{subfigure}

\newcommand\aap{{\it A\&A}}%
\newcommand\mnras{{\it MNRAS}}%
\newcommand\apj{{\it ApJ}}%
\newcommand\aj{{\it AJ}}%
\newcommand\nat{{\it Nature}}%

\title[PAnDAS M31 Globular Cluster Abundances] 
{Chemical Abundances of Seven Outer Halo M31 Globular Clusters from the Pan-Andromeda Archaeological Survey}

\author[Charli M. Sakari]   
{Charli M. Sakari$^1$}

\affiliation{$^1$Department of Astronomy, University of Washington, Seattle, WA
98195-1580, USA}

\pubyear{2016}
\volume{xxx}  
\jname{Title of your IAU Symposium}
\editors{A.C. Editor, B.D. Editor \& C.E. Editor, eds.}
\begin{document}

\maketitle

\begin{abstract}
Observations of stellar streams in M31's outer halo suggest that M31
is actively accreting several dwarf galaxies and their globular
clusters (GCs). Detailed abundances can chemically link clusters to
their birth environments, establishing whether or not a GC has
been accreted from a satellite dwarf galaxy. This talk presents the
detailed chemical abundances of seven M31 outer halo GCs
(with projected distances from M31 greater than 30 kpc), as derived
from high-resolution integrated-light spectra taken with the Hobby
Eberly Telescope. Five of these clusters were recently discovered in
the Pan-Andromeda Archaeological Survey (PAndAS)---this talk presents
the first determinations of integrated Fe, Na, Mg, Ca, Ti, Ni, Ba, and
Eu abundances for these clusters. Four of the target clusters (PA06,
PA53, PA54, and PA56) are metal-poor ($[\rm{Fe/H}] < -1.5$),
$\alpha$-enhanced (though they are possibly less alpha-enhanced than
Milky Way stars at the 1 sigma level), and show signs of star-to-star
Na and Mg variations. The other three GCs (H10, H23, and
PA17) are more metal-rich, with metallicities ranging from [Fe/H] =
-1.4 to -0.9. While H23 is chemically similar to Milky Way field
stars, Milky Way GCs, and other M31 clusters, H10 and
PA17 have moderately-low [Ca/Fe], compared to Milky Way field stars
and clusters. Additionally, PA17's high [Mg/Ca] and [Ba/Eu] ratios are
distinct from Milky Way stars, and are in better agreement with the
stars and clusters in the Large Magellanic Cloud (LMC). None of the
clusters studied here can be conclusively linked to any of the
identified streams from PAndAS; however, based on their locations,
kinematics, metallicities, and detailed abundances, the most
metal-rich PAndAS clusters H23 and PA17 may be associated with the
progenitor of the Giant Stellar Stream, H10 may be associated with the
SW Cloud, and PA53 and PA56 may be associated with the Eastern Cloud.

\end{abstract}

\firstsection 
\section{Introduction}
Globular clusters (GCs) can be used to probe the properties of their
host galaxies' field stars.  This is especially important for distant
systems whose field stars are unresolved but whose GCs can be observed
with integrated-light (IL) spectroscopy.  Detailed abundances of GCs
are useful for examining the formation and evolution of a galaxy,
particularly whether its GCs formed in a massive galaxy or were
accreted from dwarf satellites.  This process of chemical tagging is
possible with high-resolution IL spectroscopy, and can be applied to a
variety of GC systems.  This talk showcases the results from
\citet{Sakari2015}, who determined integrated abundances of seven
outer halo ($R_{\rm{proj}} > 30$ kpc) M31 GCs, five of which were
discovered in the Pan-Andromeda Archaeological Survey (PAndAS;
\citealt{McConnachie2009,Huxor2014}).  The location of these GCs in
the outer halo, their kinematics \citep{Veljanoski2014}, and/or their
projected locations along stellar streams \citep{Mackey2010} strongly
suggest an accretion scenario for many of these GCs.  Detailed
abundances can also be useful for determining the properties of the
host dwarf galaxies.

\section{Integrated Chemical Abundances of Outer Halo M31 GCs}
Spectra of the seven target GCs were obtained with the High Resolution
Spectrograph on the Hobby-Eberly Telescope, yielding a spectral
resolution of $R=30000$ (see \citealt{Sakari2015} for details).  S/N
ratios ranged from $\sim 40-140$ per resolution element.  {\it Hubble
  Space Telescope} partially resolved photometry (down to the
horizontal branch) was used to constrain the ages and metallicities of
the GC stellar populations; the final adopted ages and metallicities
were then determined spectroscopically (see
\citealt{McWB,Colucci2014,Sakari2013}).  BaSTI isochrones
\citep{BaSTIref} were adopted to model the underlying
populations. Integrated abundances were determined with the {\tt
  ILABUNDS} code \citep{McWB}.\\

\begin{figure}[b]
\begin{center}
\includegraphics[scale=0.6]{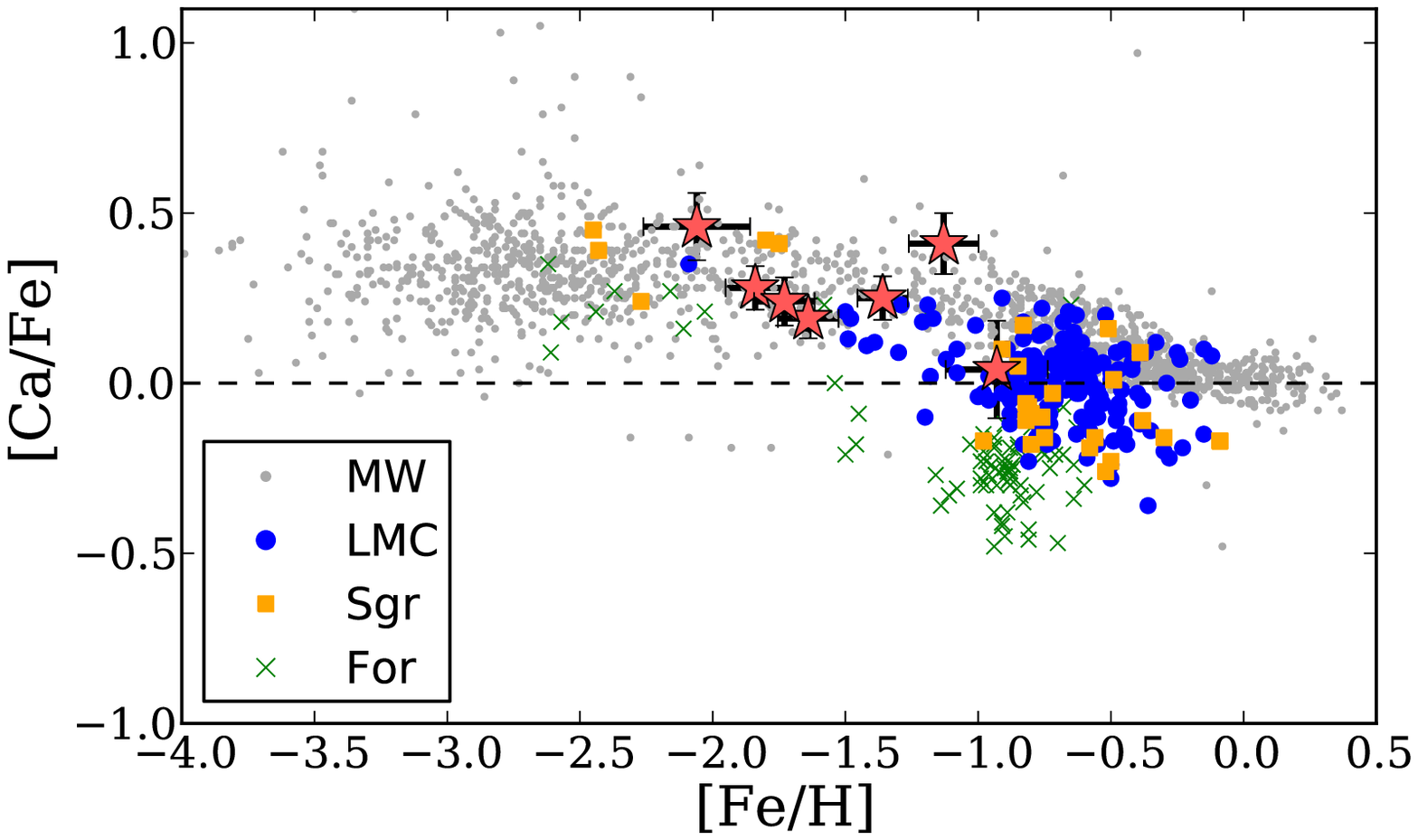}
 \caption{Comparisons of PAndAS clusters (red stars) to
MW field stars (grey; from the sources in \citealt{Venn2004}, with
supplements from \citealt{Reddy2006}) and dwarf galaxy field stars.
LMC stars are shown with blue circles \citep{Pompeia}, Sgr with orange
squares \citep{Sbordone2007,Monaco2007,CarrettaSgr,ChouSgr,McW2013},
and Fornax with green crosses
\citep{Tafelmeyer2010,Shetrone2003,Letarte2010}.  The error bars show
the random errors.  The dashed horizontal line shows the solar value.
}
   \label{fig1}
\end{center}
\end{figure}

Detailed abundances of a variety of elements (Fe, Na, Mg,
Ca, Ti, Ni, Ba, and Eu) are given in \citet{Sakari2015}.  Figure
\ref{fig1} shows integrated [Ca/Fe] versus [Fe/H] in the PAndAS GCs
compared to the abundances from individual field stars in the
Milky Way and various satellite galaxies.  Note that the GC Ca
abundances follow the chemical evolution profile of field stars that
formed in the same birth environment (e.g.,
\citealt{Pritzl2005,Sakari2013,Colucci2014,Hendricks2016}); the
integrated [Ca/Fe] is therefore a useful abundance ratio for chemical
tagging purposes.\footnote{Other
  integrated GC abundances may not follow the field stars, e.g., Na and
  Mg.  These abundances are useful for probing the nature of GCs (see
  \citealt{Sakari2013,Sakari2015,Colucci2014}).} Several of the PAndAS
GCs have low [Ca/Fe] (and other ratios, such as high [Ba/Eu])
indicating that they likely formed in dwarf galaxies that are
currently being accreted by M31.  The abundances themselves are also
useful for constraining the mass of the progenitor dwarf
galaxies---for example, PA17 (the most metal-rich target) has
abundances that are more consistent with a galaxy like the Large
Magellanic Cloud (LMC) or the Sagittarius dwarf spheroidal (Sgr) than
with the Fornax dwarf (For).

\section{The Nature of M31's Outer Halo}
These detailed abundances can be combined with kinematic and spatial
information to link GCs with specific progenitor streams and/or other
GCs.  PA17's high [Fe/H] and low [Ca/Fe] suggest that it may be
associated with the metal-rich Giant Stellar Stream (GSS; see, e.g.,
\citealt{Ibata2014})---though it does not lie along the GSS, models
suggest that the stream may have multiple wraps around M31
\citep{Fardal2013,Kirihara2016}.  The moderate-metallicity GC H10 lies
on an extension of the SW Cloud, and has similar abundances and
kinematics to its GCs \citep{Mackey2013}. H10's [Ca/Fe] implies that
the SW Cloud progenitor likely had a similar mass as Sgr. Finally, the
more metal-poor GCs PA53 and PA56 have similar abundances to each
other, plus similar kinematics to two GCs that lie along the nearby
Eastern Cloud \citep{McMonigal2016}.  The abundances of PA53 and PA56
indicate that the progenitor of the Eastern Cloud was likely similar
to the Fornax dwarf galaxy.\\

Together, the combination of resolved photometry and IL spectroscopy
has demonstrated that M31's outer halo is at least partially been
built up by the accretion of satellite dwarfs. Studies of the outer
halo field stars and the GCs have provided essential information about
the masses of the satellites that are being accreted to form M31's
outer halo.  This level of detail will be difficult to achieve in more
distant galaxies, particularly when the field stars cannot be
resolved.  When resolved photometry is no longer feasible, IL
spectroscopy of GCs remains a powerful tool for unraveling the
assembly histories for distant, unresolved systems.

\end{document}